\RequirePackage{fix-cm}
\documentclass[a4paper]{svjour3}                     
\smartqed  
\usepackage{graphicx}
\usepackage{color}
\usepackage{amssymb,amsmath}
\usepackage{epstopdf}

\begin{document}

\title{Experimental Evidence of Near-Wall Reverse Flow Events in a Zero Pressure Gradient Turbulent Boundary Layer}
\titlerunning{Near-wall flow reversal in a ZPG turbulent boundary layer}        

\author{Christian E. Willert$^1$}


\institute{C. Willert \at
            German Aerospace Center (DLR) \\
            Institute for Propulsion Technology \\
            51170 K\"oln, Germany \\
            \email{chris.willert@dlr.de}
}

\date{Submitted \today}
\maketitle

\begin{abstract}
This study reports on experimentally observed near-wall reverse flow events in a fully developed flat plate boundary layer at zero pressure gradient with Reynolds numbers between $Re_\tau = 1000$ and $Re_\tau = 2700$. The reverse flow events are captured using high magnification particle image velocimetry sequences with record lengths varying from 50,000 to 126,000 samples. Time resolved particle image sequences allow singular reverse flow events to be followed over several time steps whereas long records of nearly statistically independent samples provide a variety of single snapshots at a higher spatial resolution. The probability of occurrence lies in the range of 0.01\% to 0.1\% which matches predictions made with direct numerical simulations (DNS). The self-similar size of the reverse flow bubble is about 30-50 wall units in length and 5 wall units in height which also agrees well to DNS data provided by Lenaers et al. (ETC13, Journal of Physics: Conference Series 318 (2011) 022013).
\end{abstract}




\section{\label{intro}Introduction}
The occurence of near wall flow reversal and with it the presence of negative values of the local wall shear stress $\tau_w$ of turbulent boundary layers have been subject of debate over the past decades. Eckelmann \cite{Eckelmann:1974} postulated that near wall reverse flow was not possible and experimentalists have rarely, if at all, observed this somewhat counter-intuitive flow phenomen.
On the other hand a variety of direct numerical simulations (DNS) suggest the opposite. For DNS of zero pressure gradient turbulent boundary-layers (ZPG TBL) events of negative shear stress have been reported by Spalart and Coleman \cite{SpalartColeman:1997} and also for a turbulent channel flow by Hu et al. \cite{HuMorfeySandham:2006}. Similar observations have been made by Lenaers et al. \cite{Lenaers:2013} using simulations of turbulent channel flow up to $Re_\tau = 1000$ and for turbulent pipe flow by Khoury et al. \cite{KhourySchlatter:2014}. Cardesa et al. \cite{Cardesa:2014} also confirm the existence of areas of vanishing wall shear stress in DNS of turbulent channel flow at $Re_\tau = 934$ and $Re_\tau = 1834$ and associate these so-called critical points with large scale structures that extend up to 800 wall units downstream.

Common to the observations of the DNS data is that with increasing Reynolds number both the occurence and the magnitude of the negative axial/streamwise velocities increase.
Lenaers et al. \cite{Lenaers:2013} report reverse flow occurence of $0.01\%$ for $Re_\tau = 180$ increasing to $0.06\%$ for $Re_\tau = 1000$. In their DNS of fully turbulent channel flow Hu et al. \cite{HuMorfeySandham:2006} report a probability of negative wall shear ($\tau_w < 0$) of 0.003\% at $Re_\tau = 90$ increasing to 0.085\% at $Re_\tau = 1440$.

Due to their predicted low occurrence reverse flow phenomena have only been observed rather seldom in experiments involving ZPG wall bounded flows. To properly catch these events long records are necessary which until recently has only been possible for single point techniques, for instance through the use of laser Doppler velocimetry in a ZPG TBL as reported by Johansson \cite{Johansson:1988}. At the same time the employed measurement technique needs to provide adequate spatial resolution as the reverse flow structures observed in DNS data are both short-lived and restricted to the viscous sublayer ($O(5 y^+)$). Using the micro pillar shear stress imaging technique, Br\"{u}cker \cite{Bruecker:2015} has recently been able to visualize the areas of reverse flow on a flat plate turbulent boundary layer at $Re_\tau \approx 940$.

Flow topology can nowadays be obtained through particle image velocimetry (PIV), yet, in comparison to single point techniques, PIV is generally restricted in acquisition frequency, number of samples and measurement uncertainty. This can be partially overcome by restricting the camera field of view which allows both an increase of sample rate and sample count \cite{Willert:2015}. The following reports on PIV measurements in the near wall area of a ZPG TBL using sample counts of up to 120,000 which is shown to be sufficient to capture several instance of reverse flow events.

The PIV measurements were primarily conducted to characterize the upstream conditions for a different experiment performed further downstream within the 20\,m long test section.  Long records, some of which are temporally resolved, enable the capture of rare events such as those described in the following.

\section{\label{sec:facility}Wind tunnel facility}
The measurements were performed at the turbulent boundary layer wind tunnel at the Laboratoire de M\'{e}canique de Lille (LML). The measurement locations are located about 3.2\,m and 6.8\,m downstream of the boundary layer trip position which itself is located at the junction between the contraction nozzle and the $2\times1\,\mbox{m}^2$ rectangular test section. The tripping device consists of a 4mm rod attached to the tunnel wall followed by a 100\,mm wide strip of coarse sandpaper (roughness $\approx 25$grid). Full optical access to the 20\,m long rectangular test section is provided by large glass windows on all for sides.

Data was acquired at two free stream velocities of $U_\infty = 5\,$m/s and $U_\infty = 9\,$m/s with the wind tunnel stabilized to within $\pm0.01\,$m/s. Temperature stabilization was set at $20.0\pm0.1\,^\circ$C. Table\,\ref{tbl:tunnelparams} provides the relevant parameters of the turbulent boundary layer at the specific measurement conditions. While the friction velocity can be retrieved directly from the PIV measurements -- using the methodology described in \cite{Willert:2015} -- other parameters such as the boundary layer thickness are estimated from theory since only the lower portion of boundary layer was captured by the high resolution PIV measurements.

For the analysis of the data the following orthogonal coordinate system is adopted: the $X$-axis is aligned in streamwise direction, $Y$ is wall-normal and $Z$ is aligned in spanwise direction.

\begin{table*}
    \caption{Global parameters of the boundary layer experiments with estimated values given in parenthesis}
    \label{tbl:tunnelparams}       
    \begin{tabular}{lllllll}
    \hline\noalign{\smallskip}
    measurement location & $x_o$ & [m] \hspace{5mm} & 3.21 & 3.21 & 6.76 & 6.76 \\[2pt]
    free stream velocity & $U_\infty$ &  [m s$^{-1}$] & 5.0 & 9.0 & 5.0 & 9.0 \\[2pt]
    boundary layer thickness & $\delta$ & [mm] & (76) & (68) & (140) & (124) \\[2pt]
    wall shear rate & $\dot{\gamma} = \partial u / \partial y |_0$ & [s]$^{-1}$ & 2950 & 8300 & 2650 & 7420 \\[2pt]
    friction velocity & $u_\tau$ & [m s$^{-1}$] & 0.212 & 0.355 & 0.201 & 0.336 \\[2pt]
    Reynolds number  & $Re_x$ & $\times 10^6$ & 1.06 & 1.90 & 2.24 & 4.03 \\[2pt]
    friction Reynolds number  & $Re_\tau$ &  & 1066 & 1590 & 1842 & 2740 \\[2pt]
    wall unit  & $y^+$ & $\mu$m & 71.8 & 42.8 & 75.7 & 45.3  \\[2pt]
    \noalign{\smallskip}\hline
    \end{tabular}
\end{table*}

\section{\label{sec:piv}PIV measurement and post-processing}
Following the procedures laid out \cite{Willert:2015} only a narrow wall-normal strip was imaged by the PIV camera(s), primarily to obtain the wall-normal velocity profile and related higher order statistics.

Two camera types were used to capture different image sequences. A high-speed CMOS camera with 36GB of RAM (Dimax-S4, PCO GmbH, Germany) could capture more than 126,000 frames at 6.7\,kHz to provide continuous time records. By reducing sample rates to 1-2\,kHz statistical independence of the samples was improved while maintaining a similar sample count. Additional measurements were performed using a scientific CMOS PIV camera (Edge 5.5, PCO GmbH, Germany) which featured increased sensitivity and higher spatial resolution. This camera was operated at a double frame rate of 200\,Hz to capture long records of statistically independent samples.

The roughly 5\,mm wide measurement area was illuminated by a pair of externally modulated continuous wave lasers (Kvant Laser, Slovakia) with a combined output power of about 10\,W at a wavelength of 520\,nm. The non-collimated laser beam with a size of about $6 \times 2$\,mm$^2$ was focussed into a uniform 6\,mm wide light sheet using a cylindrical lens ($fl=200$\,mm). The resulting waist thickness was on the order of $200\,\mu$m before entering the wind tunnel glass panel from below.

Seeding was provided globally in the closed circuit wind tunnel. Consisting of an evaporated-recondensed water-glycol mixture, it was introduced in the diffuser downstream of the 20\,m long test section just upstream of the fan. The size of the aerosol droplets was estimated at $1\mu$m with a lifetime in the order of 10\,minutes.

A telephoto lens (Zeiss Apo-Tessar 300\,mm/f2.8) with a 100\,mm extension tube imaged the near wall region with a sufficiently high magnification ($m=0.44$) at working distance of $\approx 1.1$\,m from the tunnel's centerline. For the high-speed camera with a pixel size of $11\,\mu$m this results in a magnification $25.4\,\mu$m in object space. For the sCMOS camera, with $6.5\,\mu$m pixel pitch, the spatial resolution improves to $14.1\,\mu$m per pixel. To make use of nearly the full aperture of the objective lens the optical axis was inclined about $1.5^\circ$ with respect to the tunnel wall.

\begin{table*}
    \caption{PIV parameters of the boundary layer experiments }
    \label{tbl:pivparams}       
    \begin{tabular}{lllll}
    \hline\noalign{\smallskip}
    {Camera model} &  &  \hspace{15mm} & \textbf{PCO Dimax-S4} \hspace{5mm}& \textbf{PCO Edge 5.5} \\[2pt]
    pixel size &  &  [$\mu\mbox{m}^2$] & $11.0\times 11.0$ & $6.5\times 6.5$ \\[2pt]
    magnification & $m$ &  [$\mu$m pixel$^{-1}$] & 25.4 & 14.1 \\[2pt]
    image size &  $H \times W$ &  [pixel] & $200\times 1008$ & $200\times 2560$     \\[2pt]
    field of view & $w \times h$ & [$\mbox{mm}^2$] & $5.08\times 25.6$ & $2.82\times 36.1$  \\[2pt]
    pulse separation & at 5 m/s & [$\mu$s] & 150 & 100 \\
      & at 9 m/s & [$\mu$s] & 100 & 65 \\[2pt]
    \noalign{\smallskip}\hline
    \end{tabular}
\end{table*}

The acquired data was processed using a conventional 2-C PIV processing package featuring a coarse-to-fine resolution pyramid with intermediate image deformation (PIVview2C, PIVTEC GmbH, Germany).
To obtain reliable mean velocity data and statistics within close proximity to the wall a high aspect ratio image sampling window of 64\,pixels width and 8\,pixels height was chosen. This corresponds to $1.63\times0.20\,\mbox{mm}^2$ for the high-speed camera and $0.90\times0.11\,\mbox{mm}^2$ for the sCMOS camera. For the latter the sample has an effective size of $12.6 x^+\times 1.57 y^+$ at $U_\infty = 5$\,m/s increasing to $21.1 x^+\times 2.64 y^+$ at $U_\infty = 9$\,m/s. The sample overlap was set at 75\%. Estimates of the mean and unsteady wall shear rate $\dot{\gamma}$ were obtained using a single-line cross-correlation approach (i.e. the sampling window only has a wall-normal size of one pixel).

\section{\label{sec:piv}Data analysis}
\begin{figure}[tb]
    \includegraphics[width=0.49\textwidth]{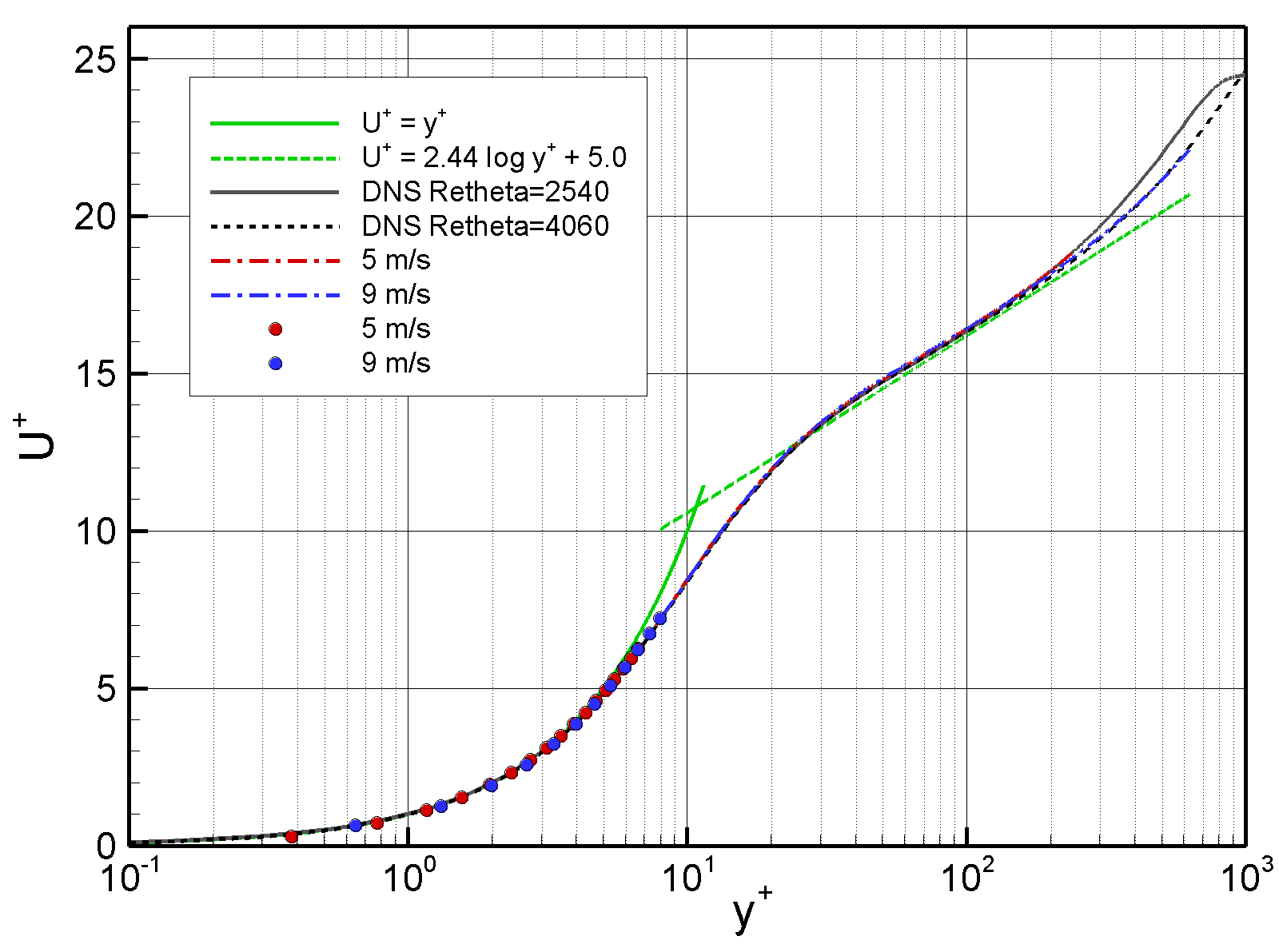}
    \includegraphics[width=0.49\textwidth]{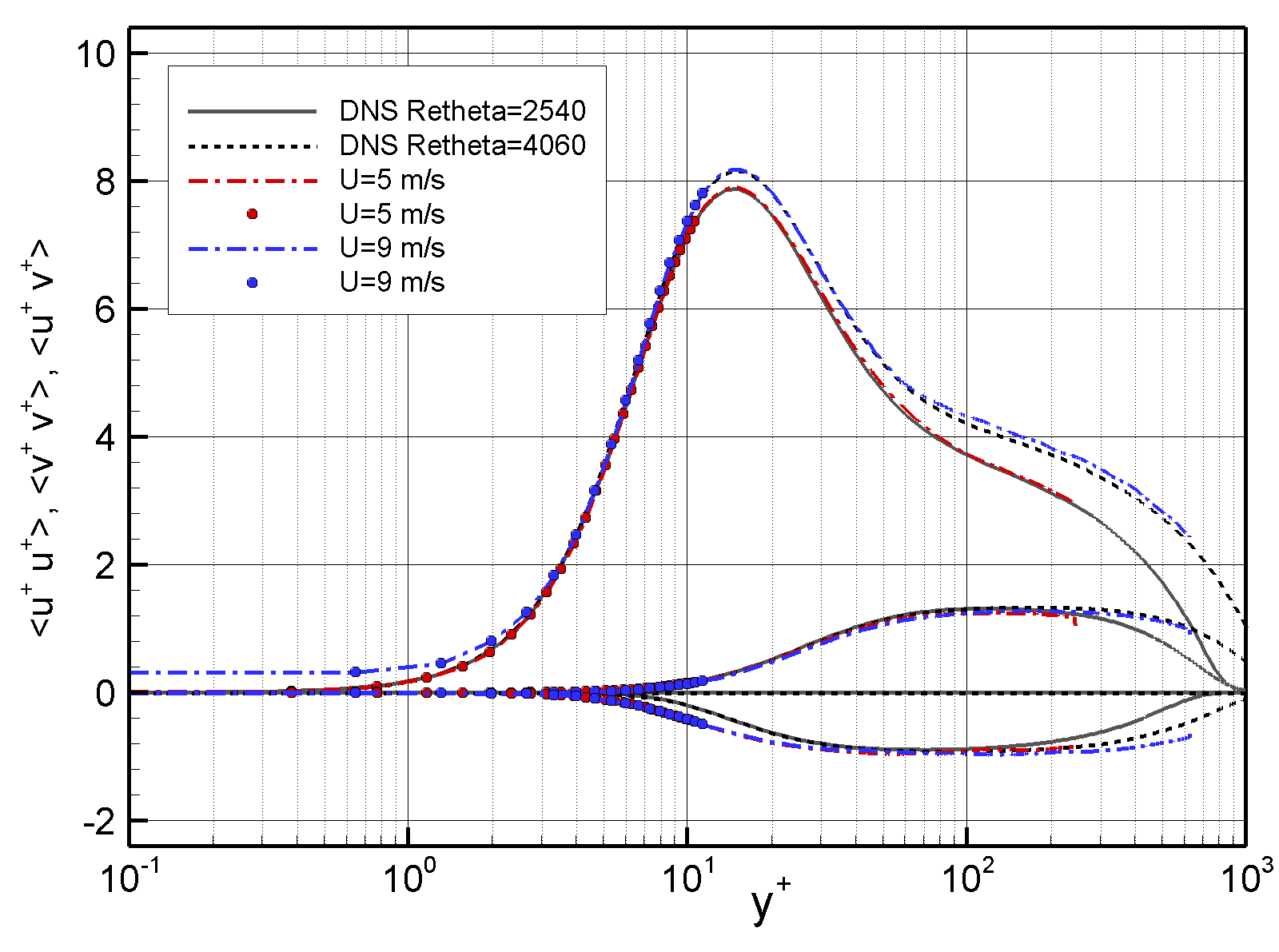}
    \caption{Profiles of the mean streamwise velocity (left) and Reynolds stresses for two different free stream velocities obtained $x=3.2$\,m downstream of the tripping device scaled with inner variables. Symbols only shown for $y^+ < 10$ for clarity.}
    \label{fig:normal_profiles}
\end{figure}

To verify that the investigated flow is representative for a ZPG TBL the following will describe some relevant statistics retrieved from the processed data sets. The measured data is normalized with inner variables using the traditional viscous scaling for velocity $u^+_i (= u_i/u_\tau)$ and  length $l^+_i (= l_i\,u_\tau/\nu)$. In this sense the mean velocity profiles for $u$ at position $x=3.2$\,m for both Reynolds numbers are shown in Fig.\,\ref{fig:normal_profiles}, left. The corresponding variances $<u'u'>,\,<v'v'>$ and covariances $<u'v'>$, are provided in Fig.\,\ref{fig:normal_profiles}, right. Both plots also contain reference data from a DNS of a ZPG TBL provided by Schlatter et al. \cite{Schlatter:2009,Schlatter:2010}. For the most part, the agreement between experiment and simulation is very good (the lines coincide). Due to the limited field of view, the outer region of the boundary is not captured. Discrepancies can be observed very to the wall for the high-Re case, stemming from the loss of resolution proportional to the reduction in viscous length scales. The covariance $<u'v'>$ (lower line grouping in Fig.\,\ref{fig:normal_profiles}, right) does exhibit some differences in both the buffer layer (around $10y^+$) and outer region ($>200y^+$) for which there currently is no obvious explanation.

\begin{figure}[tb]
    \includegraphics[width=0.49\textwidth]{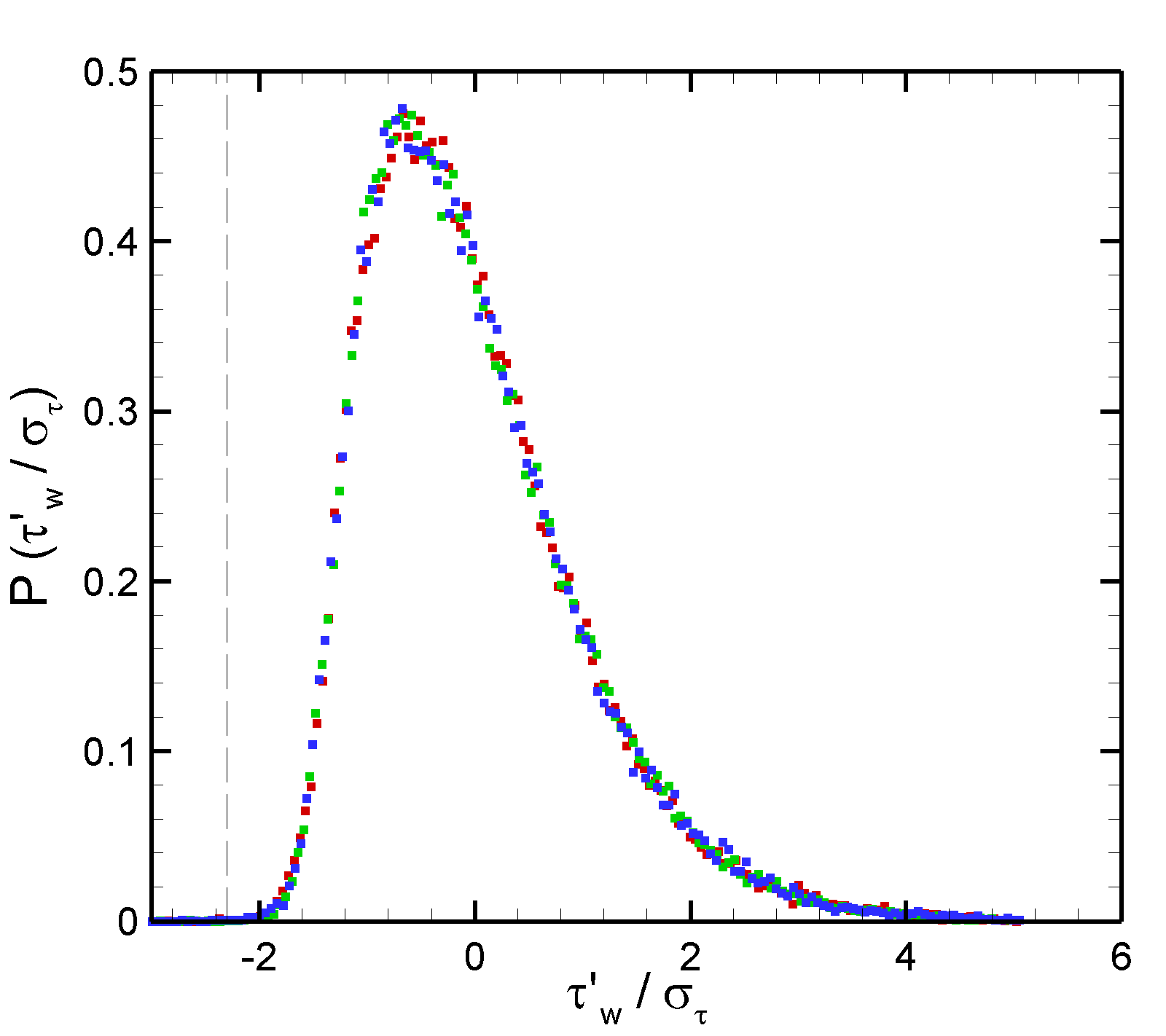}
    \includegraphics[width=0.49\textwidth]{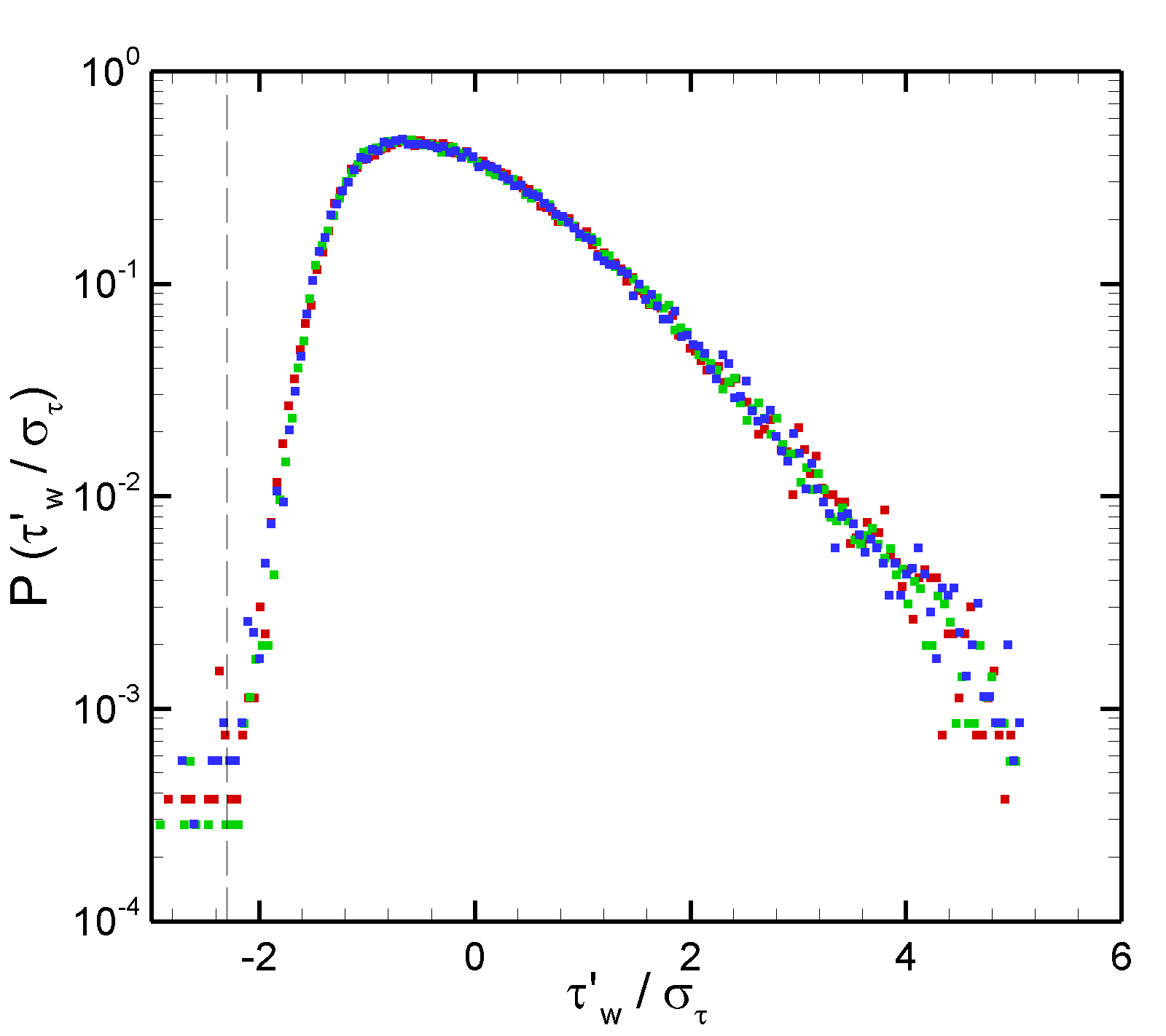}
    \caption{Probability density functions of wall shear rate $\dot{\gamma} = \partial u / \partial y$ for different data sets obtained at $Re_\tau = 1590$. Vertical line at -2.3 represents $\dot{\gamma} = 0$}
    \label{fig:wallshear_pdf}
\end{figure}

\begin{figure}[tb]
    \includegraphics[width=0.6\textwidth]{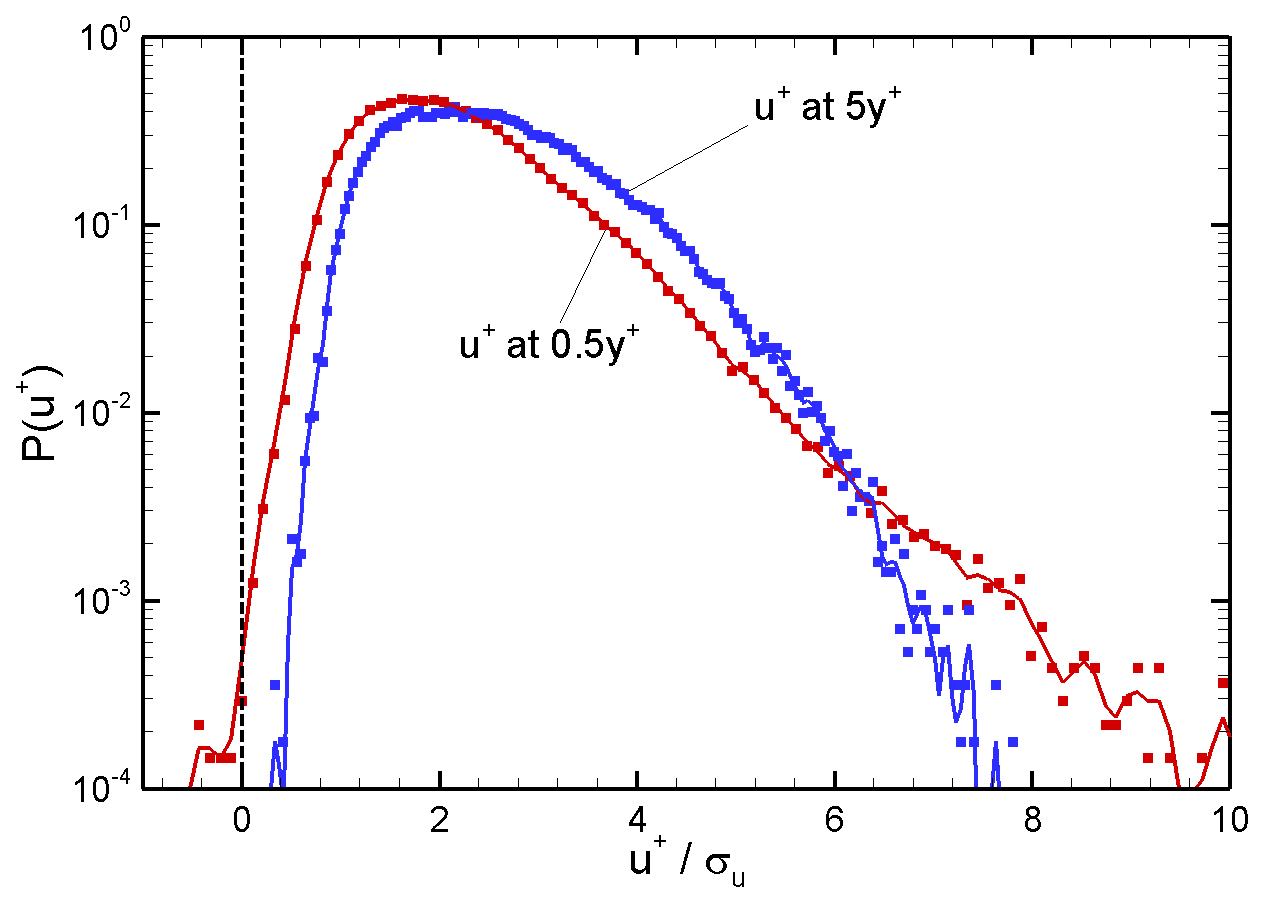}
    \caption{Probability density functions of velocity $u^+$ at wall distances of $0.5 y^+$ and $5 y^+$ for $Re_\tau = 1070$}
    \label{fig:uplus_pdf}
\end{figure}

Making use of the high spatial resolution near the wall, both the mean and unsteady wall shear rate $\dot{\gamma} = \partial u / \partial y$ and with it the corresponding wall shear stress $\tau_w = \mu\, \partial u / \partial y$ can be directly estimated from the velocity gradient at the wall. The reader is referred to \cite{Willert:2015} for details on the processing scheme. Fig.\,\ref{fig:wallshear_pdf} provides representative probability density functions (PDF) of the wall shear stress $\tau_w$ for several different sequences and show good agreement with data published in literature (e.g \cite{GrosseSchroeder:2009,HuMorfeySandham:2006,Keirsbulck:2012,Lenaers:2013,Miyagi:2000}.
The vertical line at -2.3 marks the position for $\tau_w = 0$. When plotted in log-linear form the PDF of the shear stress exhibits several instances of negative shear stress with a probability of less than $0.01\%$. Since the wall shear stress is directly related to the near-wall velocity, the PDFs of the streamwise velocity $u$ at wall distances of $1y^+$ and $5y^+$ are provided in Fig.\,\ref{fig:uplus_pdf}. Here it can be observed that the reverse flow seems to only appear very close to the wall while it is practically absent outside of the viscous sublayer for $y^+ < 5$.

\begin{figure}[tb]
    \includegraphics[width=\textwidth]{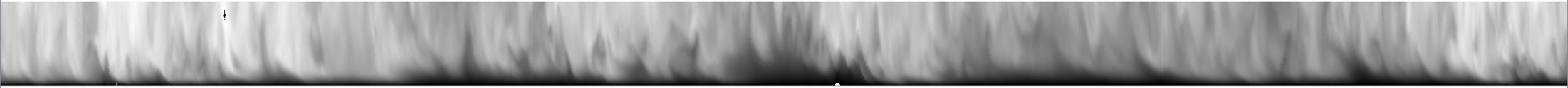}\\[2mm]
    \includegraphics[width=0.06\textwidth]{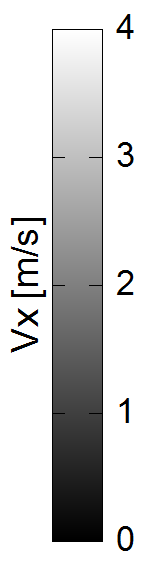}
    \hspace{0.01\textwidth}
    \includegraphics[width=0.9\textwidth]{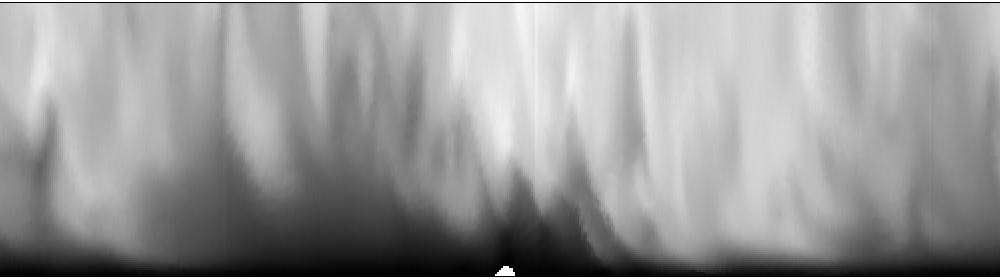}
    \caption{Time trace of streamwise velocity $u'$ at $U_\infty = 5$\,m/s  ($Re_\tau = 1840$) covering 0.9\,s (top) and 0.075\,s (500 samples, bottom). The white region near the center indicates a single reverse flow event. Vertical axis represents wall distance $0 < y < 95^+$ ($\approx 7$\,mm)}
    \label{fig:vx_vs_time}
\end{figure}

At this point it should be noted that the data in the tails of the PDFs have an increasing likelihood of being affected by measurements errors (outliers) rather than representing reliable measurements. Therefore the underlying data sets require separate verification to determine a given datum's validity which, taking into account the rather low probability of less than $0.01\%$, is feasible through visual inspection. This can be achieved through velocity-vs-time plots such as shown in Fig.\,\ref{fig:vx_vs_time}. This image is compiled by extracting a single column of data from each PIV data set and placing the columns side-by-side such that the resulting image has a width of up to 126,000 pixels, depending on the length of the complete sequence. Therefore each line in the image represents the velocity record for a given wall distance.

Reverse flow events can be easily detected by highlighting negative velocities in images such as Fig.\,\ref{fig:vx_vs_time} and retrieving the corresponding single data sets from the sequence for closer inspection. One such event is the white spot near the middle of the bottom edge of Fig.\,\ref{fig:vx_vs_time}. The spot has a size of about 6 pixels height and 12 pixels width, the former giving an indication on the vertical height of the reverse flow bubble (i.e. about $300\mu$m given a grid step size of 2 pixel). In the present sequence of 126,000 images only two such reverse flow events can be detected. Taking into account the duration of about 10-15 samples per event results in a probability of about 0.01\%.

\begin{figure}[tb]
    \includegraphics[width=0.15\textwidth]{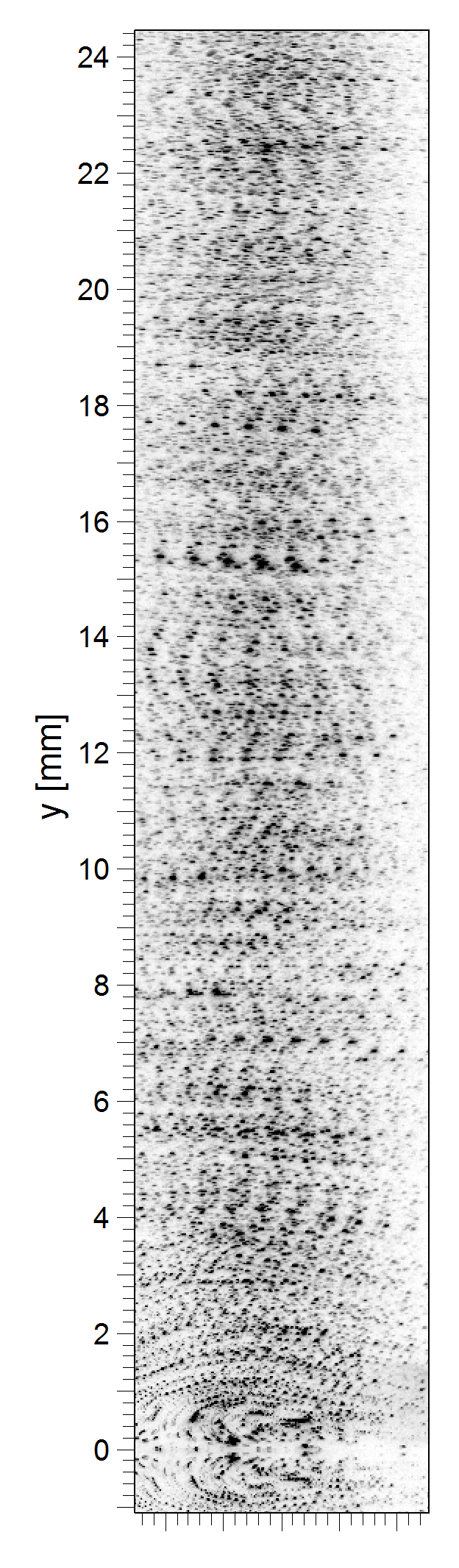}
    \includegraphics[height=0.5\textwidth]{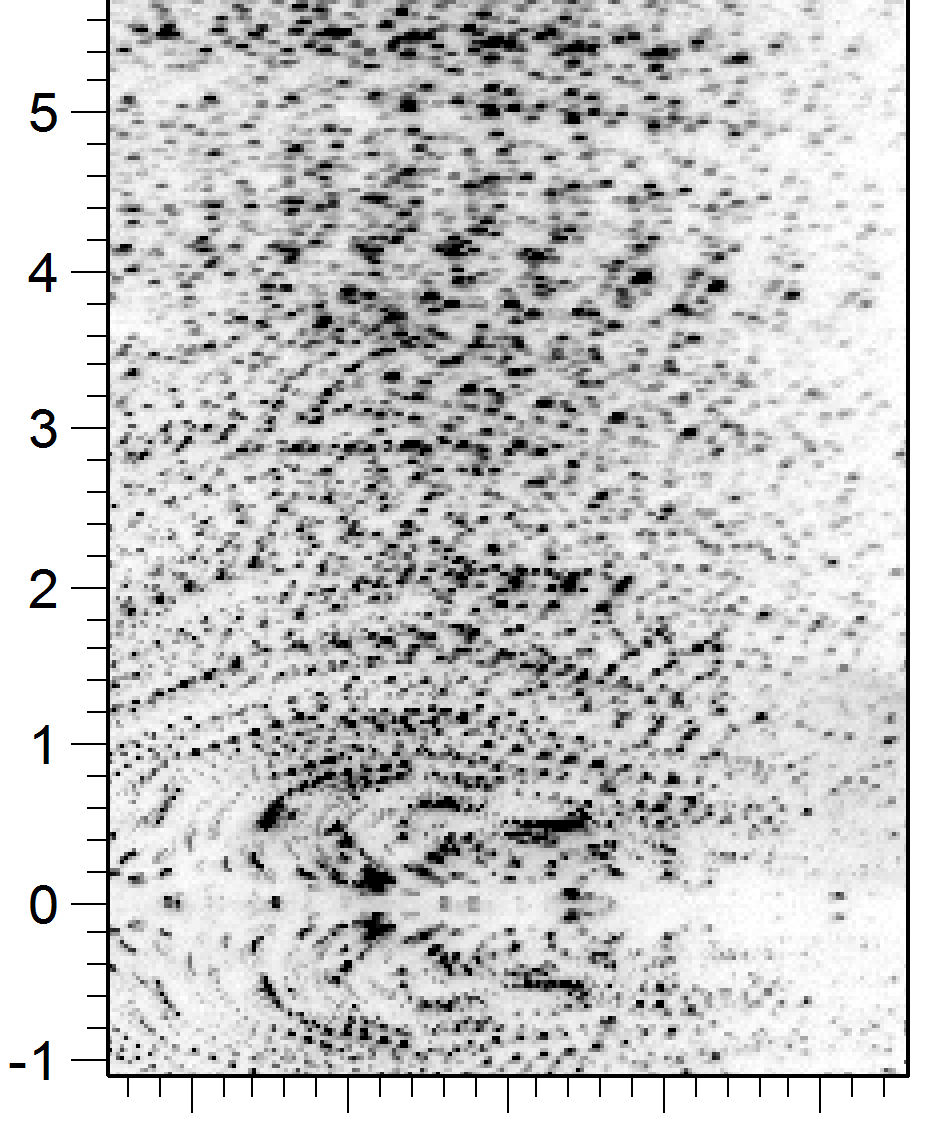}
    \caption{Multi-exposed images showing a near wall separation event in a ZPG TBL at $Re_\tau = 1840$} ($x = 6.8\,$m, $U_\infty = 5$\,m/s). Full image view at left with detail on left. The tunnel glass wall is located at $y=0$.
    \label{fig:ptv_tracks}
\end{figure}

\begin{figure}[tb]
    \hspace{0.35\textwidth}\includegraphics[width=0.25\textwidth]{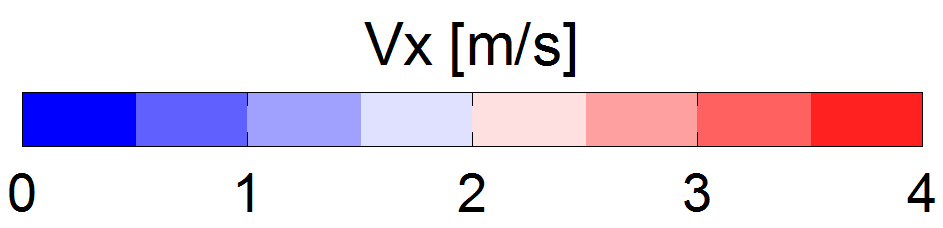}\\
    \includegraphics[height=0.2\textwidth]{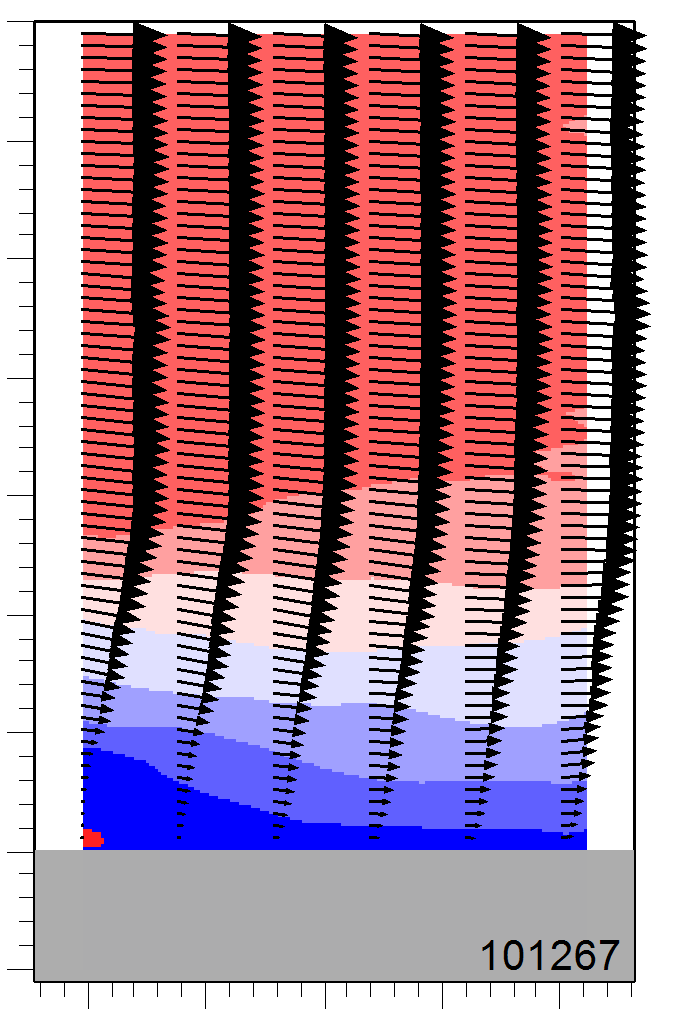}
    \includegraphics[height=0.2\textwidth]{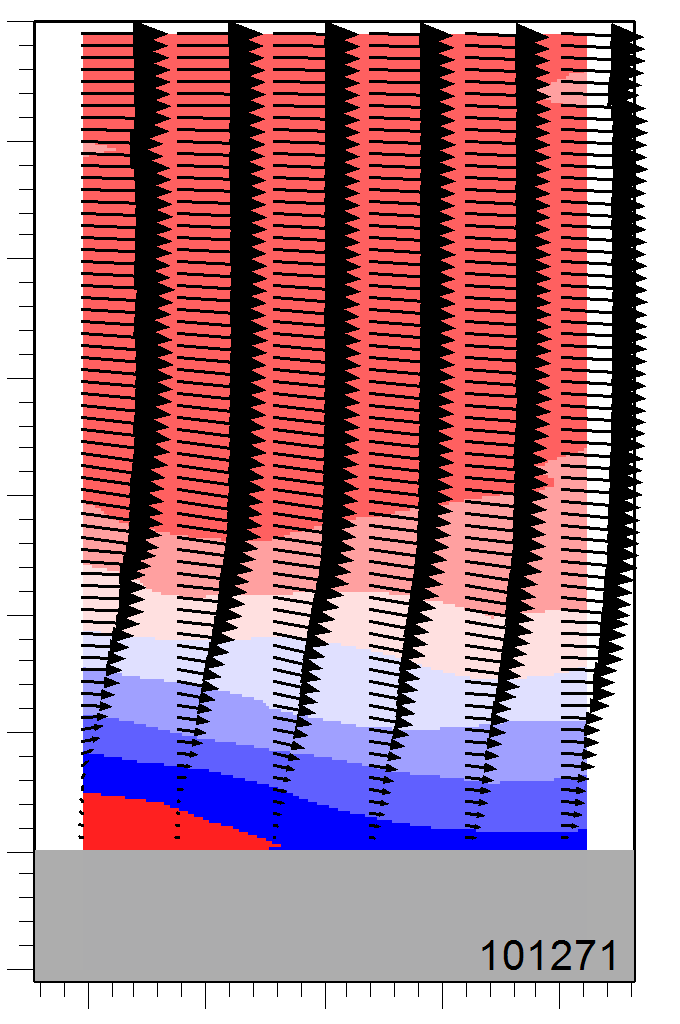}
    \includegraphics[height=0.2\textwidth]{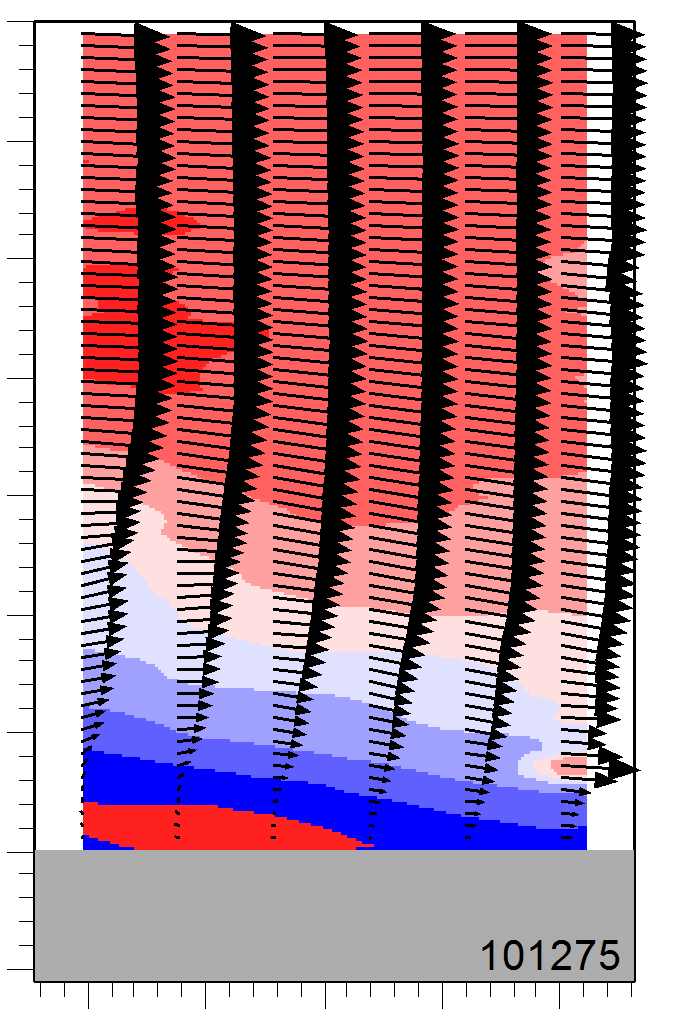}
    \includegraphics[height=0.2\textwidth]{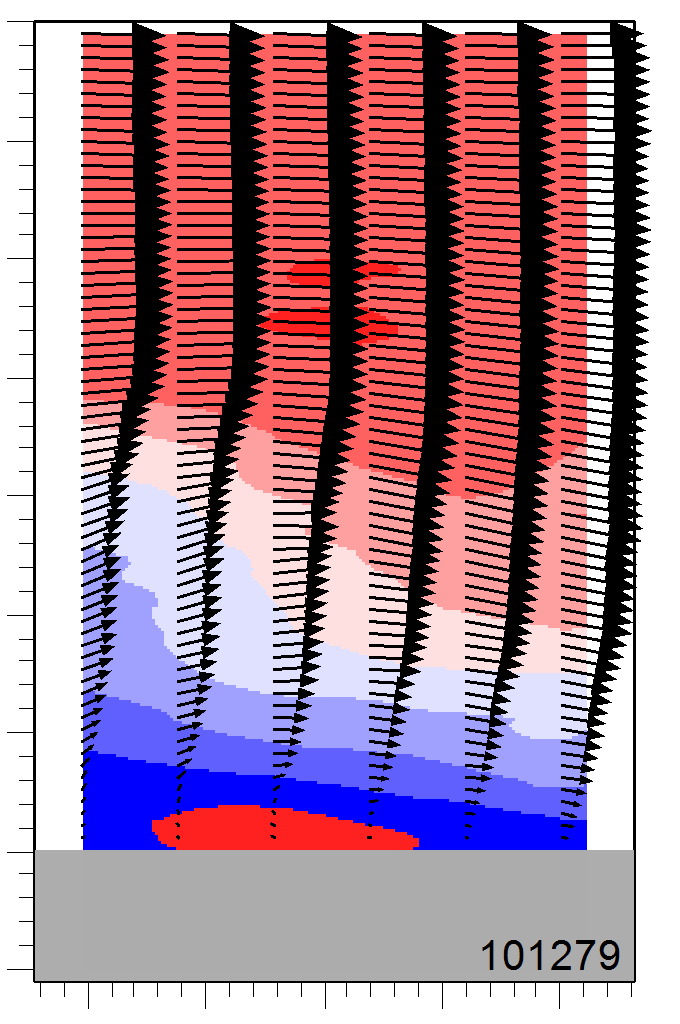}
    \includegraphics[height=0.2\textwidth]{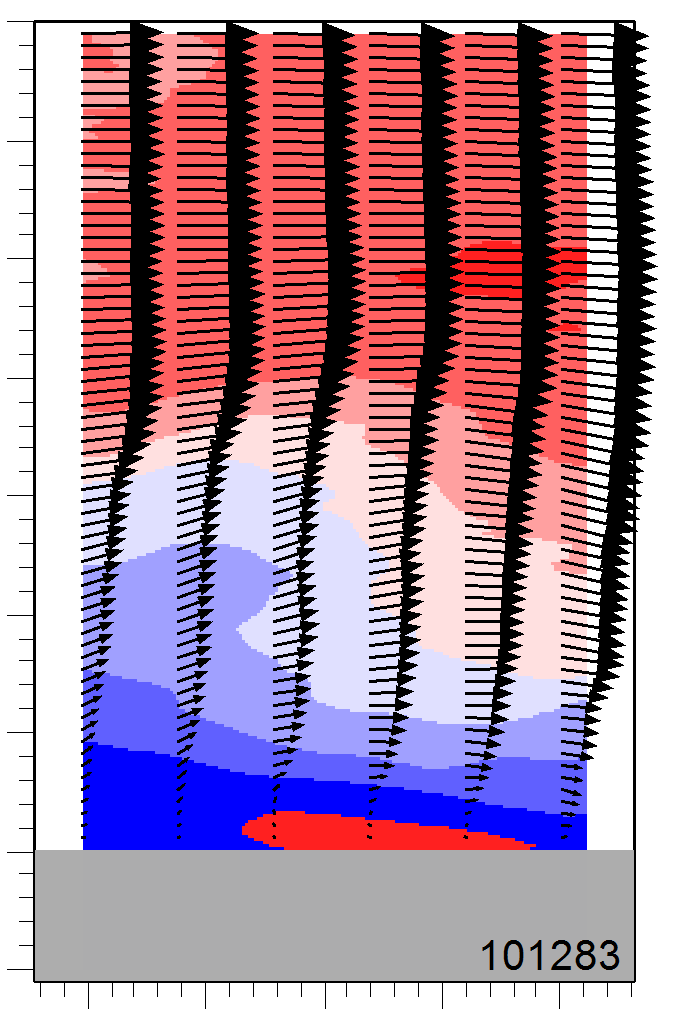}
    \includegraphics[height=0.2\textwidth]{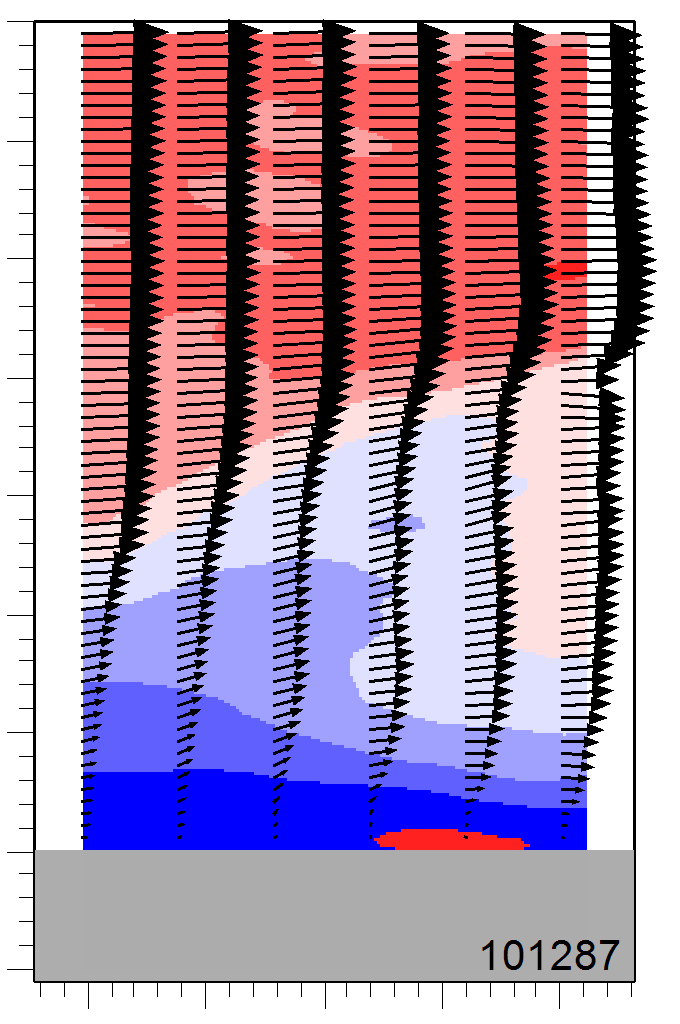}
    \includegraphics[height=0.2\textwidth]{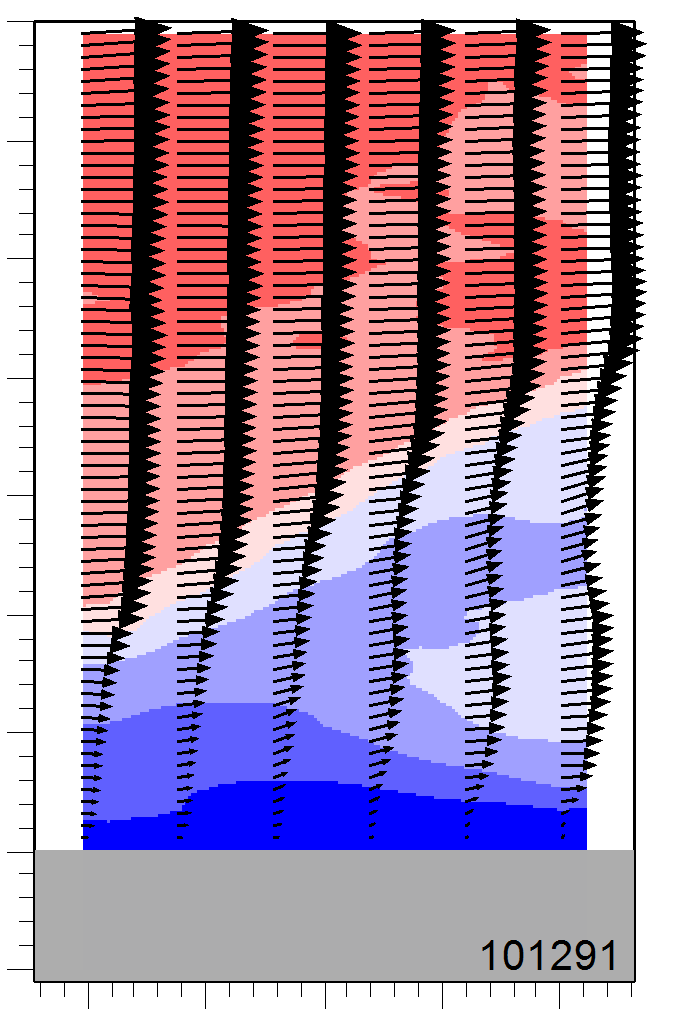}
    \caption{Sequence showing the evolution of a flow reversal event (red area) in a ZPG TBL at ($Re_\tau = 1840$). Red patch near the wall indicates reversed streamwise velocity. Temporal separation between frames is $600\,\mu$s. PIV sampling window of $32x8$\,pixels with 75\% overlap, vectors down-sampled $4\times$ horizontally, $2\times$ vertically}
    \label{fig:separation_sequence}
\end{figure}

\begin{figure}[tb]
    \includegraphics[width=0.60\textwidth]{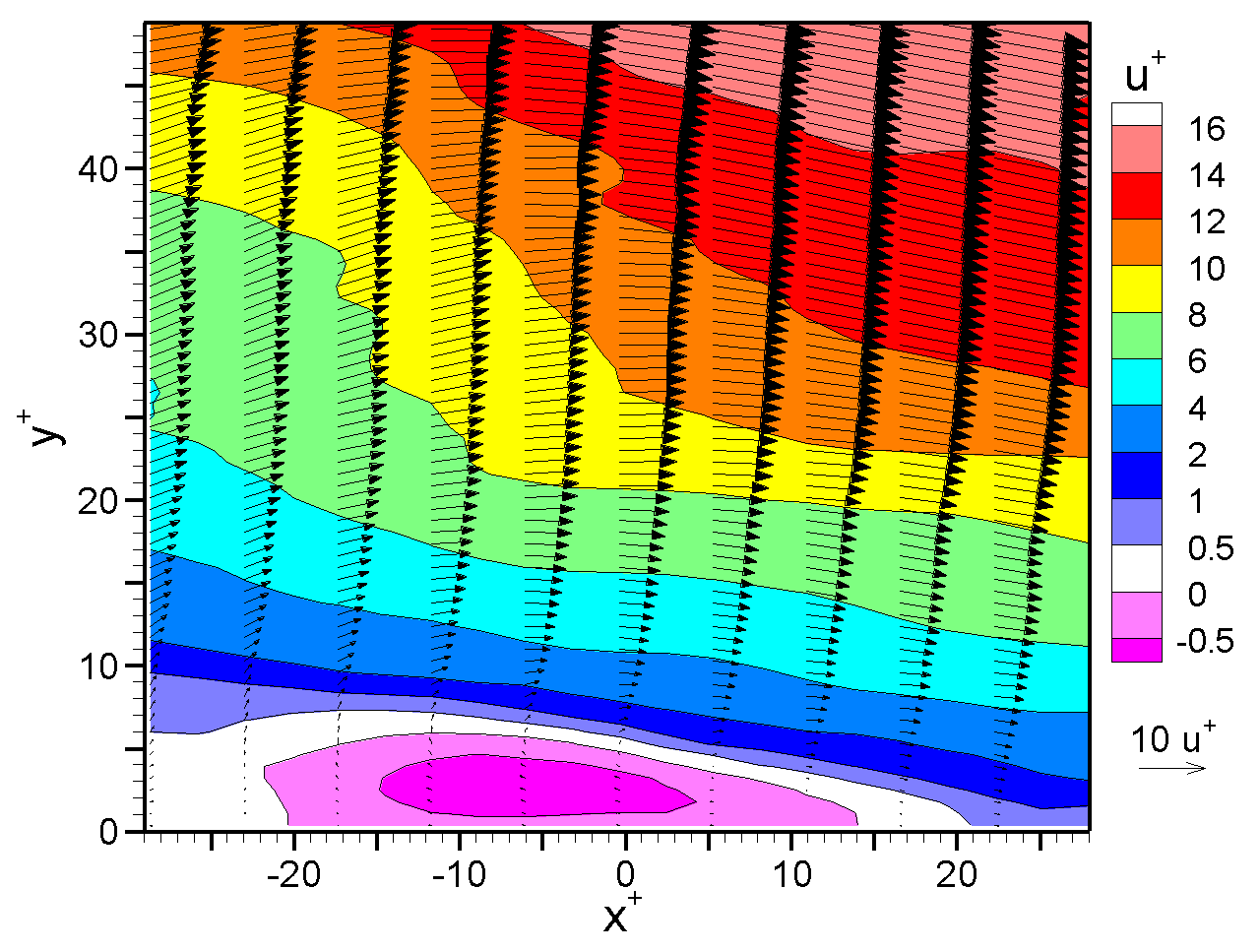}
    \caption{Detail view of center frame of Fig.\,\protect\ref{fig:separation_sequence} shown in viscous scaled units (vectors down-sampled $2\times$ horizontally)}
    \label{fig:detail_sepbubble}
\end{figure}

\begin{figure}[tb]
    \hspace{0.35\textwidth}\includegraphics[width=0.25\textwidth]{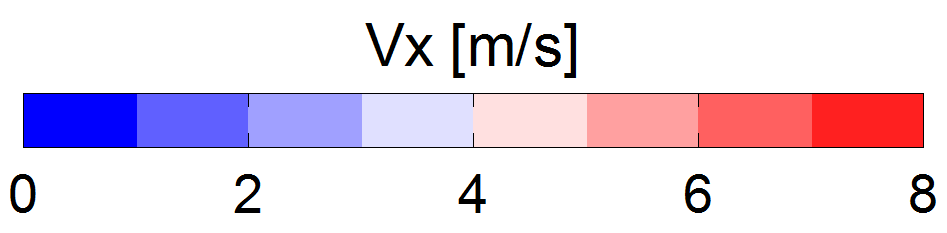}\\
    \includegraphics[height=0.22\textwidth]{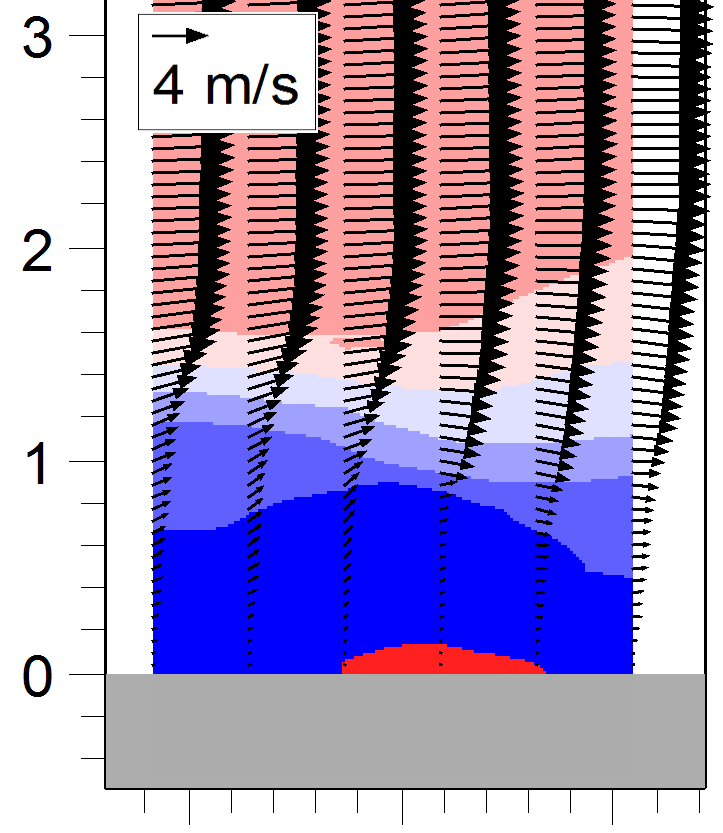}
    \includegraphics[height=0.22\textwidth]{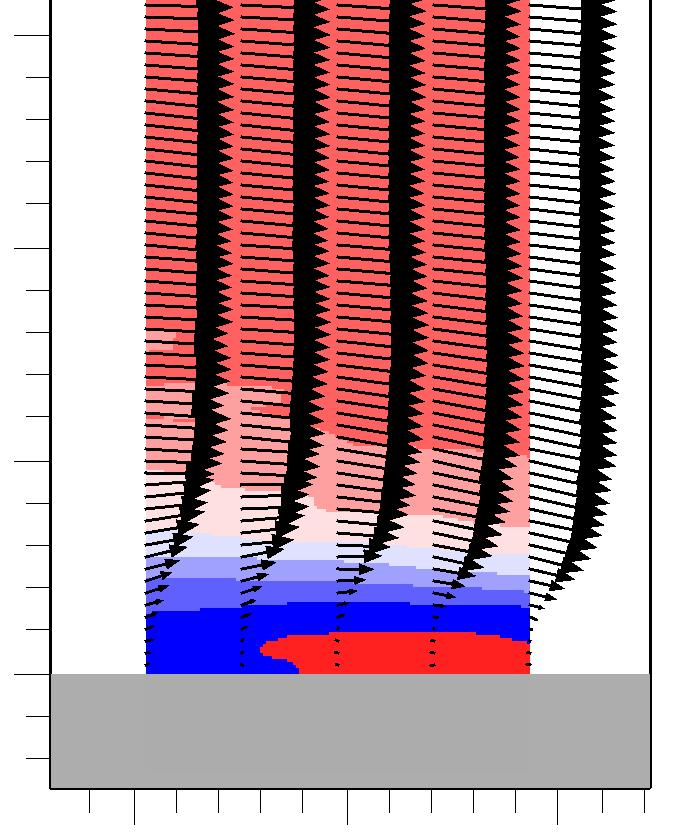}
    \includegraphics[height=0.22\textwidth]{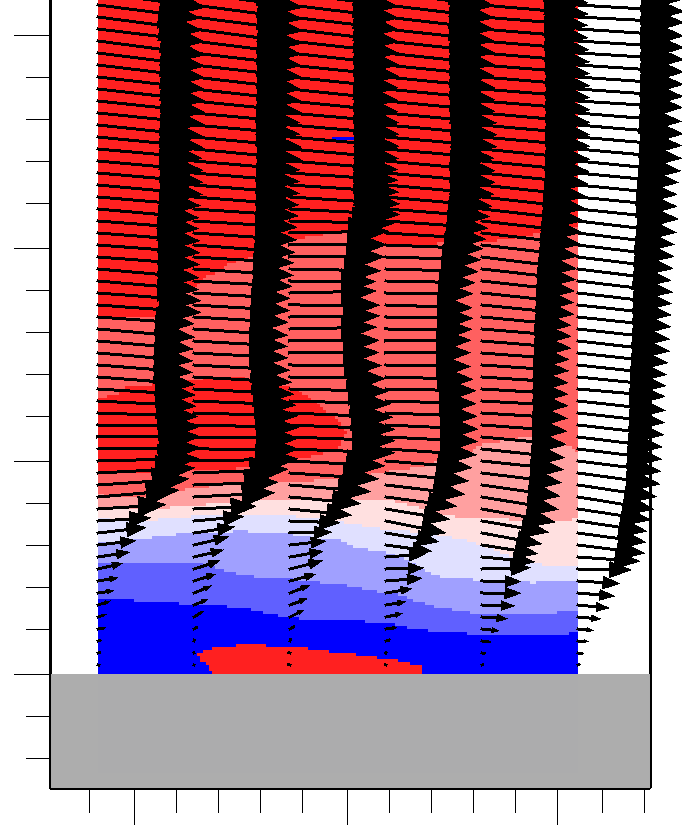}
    \includegraphics[height=0.22\textwidth]{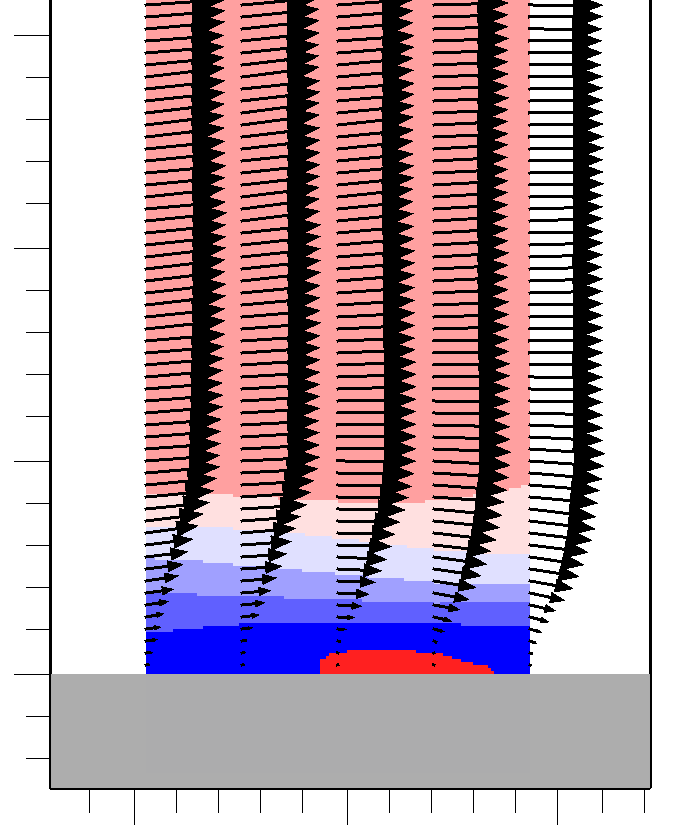}
    \includegraphics[height=0.22\textwidth]{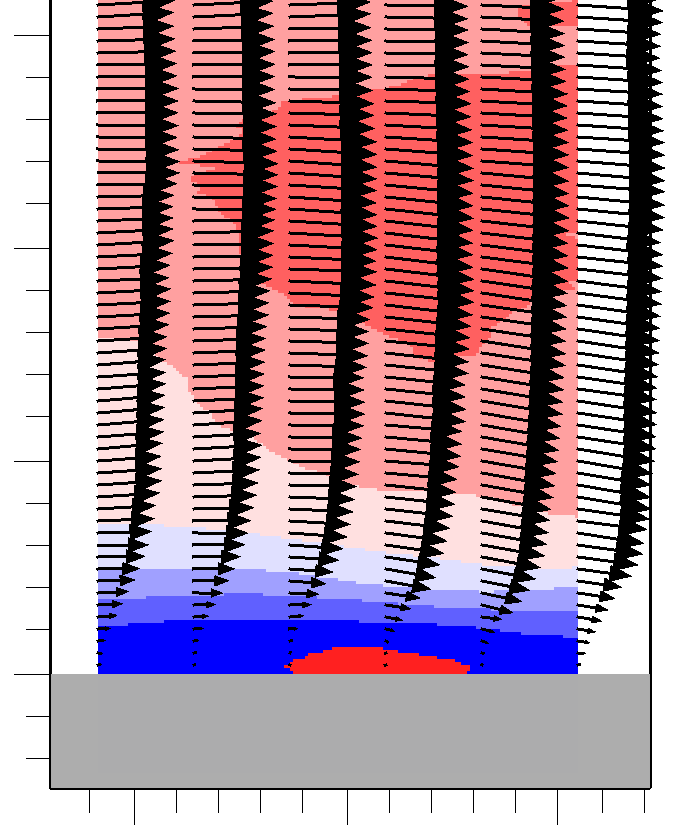}
    \caption{Various instances of near wall reverse flow indicated by red patches ($Re_\tau = 1590$). }
    \label{fig:separation_bubbles}
\end{figure}

As the image sequences are temporally well resolved, the evolution of a specific reverse flow event can be observed within the narrow field of view. In this sense the particle tracks, compiled through the summation of several images, visualize the shape of the flow structure (Fig.\,\ref{fig:ptv_tracks}). The passage of the reverse flow region through the field of view is further visualized by a sequence of velocity fields in Fig.\,\ref{fig:separation_sequence} for which only every fourth frame is shown. The red patch is this sequence defines the zone of negative streamwise velocity $u$.
A magnified view of a single data set from Fig.\,\ref{fig:separation_sequence} is provided in Fig.\,\ref{fig:detail_sepbubble}. Finally snapshots of several different "separation bubbles" are shown in Fig.\,\ref{fig:separation_bubbles} for a free stream velocity of $U_\infty = 9$\,m/s at measurement location $x=3.2$\,m.

\section{\label{sec:results}Results and discussion}
In total eight sequences at two measurement locations and two free stream velocities were investigated for the presence of reverse flow. While some sequences were affected by low seeding density and limited spatial resolution, all sequences showed clear evidence of reverse flow in the form of particles moving upstream for a certain duration. The appearance of the flow features is on the order of $0.01\%$.

In most cases the reverse flow region has a vertical dimension of $5y^+$ and a length of about $30-40x^+$ which corresponds to the DNS provided by Lenaers et al \cite{Lenaers:2013} (e.g. see Fig.\,9 in their publication). Time-resolved sequences show that the "separation bubble" traverses downstream through the field of view at a convection speed $U_c$ of about $0.1\,U_\infty$ or $U_c / u_\tau \approx 2.5$  (estimated from Fig.\,\ref{fig:separation_sequence}). While not captured through the present measurements, the corresponding DNS indicate that the structures have a spanwise dimension on the order of $30z^+$, that is, their $xz$ shape is roughly circular.

Due to their rather limited number in comparison to the number of available samples a more detailed statistical analysis is not reasonable. Furthermore, the boundary conditions of the wind tunnel have yet to be fully characterized. While the wind tunnel test section generally has a slight favorable pressure gradient (personal communication), a model was installed at $x>10$\,m further downstream of the present measurement locations. The presence of the model will undoubtedly have some upstream influence on the local pressure field which in turn can affect the flow statistics and with it the probability of flow reversal events. At the very least, the present investigation shows that the topology of low-probability reverse flow events can indeed be captured using sufficiently long PIV data sets.

\section{\label{sec:concl}Summary and outlook}
Through analysis of long PIV data sequences rare events such as small-scale near wall flow reversal could be documented at two measurement locations and two free stream velocities of the ZPG TBL. Both the probabilty of occurence as well as the shape of the observed reverse flow structure structures agree with previous DNS by Lenaers et al \cite{Lenaers:2013}. In the present measurement configuration the spanwise extension of the structures could not be measured. Multiple-camera (photogrammetric), time-resolved techniques such as tomographic PIV \cite{Scarano:2013} or 3-D PTV \cite{SchanzSchroeder:2014}, or digital holography \cite{ShengKatz:2008} are ideal candidates to capture the fully resolved velocity field of the small separation bubble.

The present data sets also exhibit strong wall-normal velocity events very close to the wall but have not been studied in detail in the present study.

\section*{Acknowledgments}
The author would like to thank his colleague J. Klinner and the staff and faculty of the LML, in particular Christophe Cuvier, for their assistance during the wind tunnel PIV measurements. Special thanks go to both the University of the Armed Forces Munich and Monash University (Melbourne) for providing their high speed cameras for the measurements.
Finally, the EU funded project EuHIT (http://www.euhit.org) is gratefully acknowledged for financial support of the measurement campaign at LML in May 2015.


\bibliographystyle{spmpsci}
\bibliography{HSPIV_TurbBL_lit}

\begin{thebibliography}{10}
\providecommand{\url}[1]{{#1}}
\providecommand{\urlprefix}{URL }
\expandafter\ifx\csname urlstyle\endcsname\relax
  \providecommand{\doi}[1]{DOI~\discretionary{}{}{}#1}\else
  \providecommand{\doi}{DOI~\discretionary{}{}{}\begingroup
  \urlstyle{rm}\Url}\fi

\bibitem{Bruecker:2015}
Br\"{u}cker, C.: Evidence of rare backflow and skin-friction critical points in
  near-wall turbulence using micropillar imaging.
\newblock Physics of Fluids \textbf{27}(3), 031705 (2015).
\newblock \doi{http://dx.doi.org/10.1063/1.4916768}

\bibitem{Cardesa:2014}
Cardesa, J.I., Monty, J.P., Soria, J., Chong, M.S.: Skin-friction critical
  points in wall-bounded flows.
\newblock Journal of Physics: Conference Series \textbf{506}(1), 012,009 (2014)

\bibitem{Eckelmann:1974}
Eckelmann, H.: The structure of the viscous sublayer and the adjacent wall
  region in a turbulent channel flow.
\newblock Journal of Fluid Mechanics \textbf{65}, 439--459 (1974).
\newblock \doi{10.1017/S0022112074001479}

\bibitem{GrosseSchroeder:2009}
Gro{{\ss}}e, S., Schr{\"{o}}der, W.: High reynolds number turbulent wind tunnel
  boundary layer wall-shear stress sensor.
\newblock Journal of Turbulence \textbf{10}(14), 1--12 (2009).
\newblock \doi{10.1080/14685240902953798}

\bibitem{HuMorfeySandham:2006}
Hu, Z., Morfey, C.L., Sandham, N.D.: Wall pressure and shear stress spectra
  from direct simulations of channel flow.
\newblock AIAA Journal \textbf{44}(7), 1541--1549 (2006)

\bibitem{Johansson:1988}
Johansson, G.: An experimental study of the structure of a flat plate turbulent
  boundary layer, using laser-doppler velocimetry.
\newblock Ph.D. thesis, Chalmers University of Technology, G{\"{o}}teborg,
  Sweden (1988)

\bibitem{Keirsbulck:2012}
Keirsbulck, L., Labraga, L., el~Hak, M.G.: Statistical properties of wall shear
  stress fluctuations in turbulent channel flows.
\newblock International Journal of Heat and Fluid Flow \textbf{37}(0), 1 -- 8
  (2012).
\newblock \doi{{10.1016/j.ijheatfluidflow.2012.04.004}}

\bibitem{KhourySchlatter:2014}
Khoury, G.K.E., Schlatter, P., Brethouwer, G., Johansson, A.V.: Turbulent pipe
  flow: Statistics, {\it{re}}-dependence, structures and similarities with
  channel and boundary layer flows.
\newblock Journal of Physics: Conference Series \textbf{506}(1), 012,010
  (2014).
\newblock \doi{10.1088/1742-6596/506/1/012010}

\bibitem{Lenaers:2013}
Lenaers, P., Li, Q., Brethouwer, G., Schlatter, P., {\"{O}}rl{\"{u}}, R.:
  Negative streamwise velocities and other rare events near the wall in
  turbulent flows.
\newblock Journal of Physics: Conference Series \textbf{318}(2), 022,013
  (2011).
\newblock \doi{10.1088/1742-6596/318/2/022013}

\bibitem{Miyagi:2000}
Miyagi, N., Kimura, M., Shoji, H., Saima, A., Ho, C.M., Tung, S., Tai, Y.C.:
  Statistical analysis on wall shear stress of turbulent boundary layer in a
  channel flow using micro-shear stress imager.
\newblock International Journal of Heat and Fluid Flow \textbf{21}, 576--581
  (2000)

\bibitem{Scarano:2013}
Scarano, F.: Tomographic {PIV}: principles and practice.
\newblock Measurement Science and Technology \textbf{24}(1), 012001 (2013).
\newblock \doi{10.1088/0957-0233/24/1/012001}

\bibitem{SchanzSchroeder:2014}
Schanz, D., Schr{\"{o}}der, A., Gesemann, S.: {'Shake The Box'} - a {4D PTV}
  algorithm: Accurate and ghostless reconstruction of lagrangian tracks in
  densely seeded flows.
\newblock In: 17th International Symposium on Applications of Laser Techniques
  to Fluid Mechanics. Lisbon, Portugal (2014)

\bibitem{Schlatter:2010}
Schlatter, P., {\"{O}}rl{\"{u}}, R.: Assessment of direct numerical simulation
  data of turbulent boundary layers.
\newblock Journal of Fluid Mechanics \textbf{659}, 116--126 (2010).
\newblock \doi{{10.1017/S0022112010003113}}

\bibitem{Schlatter:2009}
Schlatter, P., {\"{O}}rl{\"{u}}, R., Li, Q., Brethouwer, G., Fransson, J.H.M.,
  Johansson, A.V., Alfredsson, P.H., Henningson, D.S.: Turbulent boundary
  layers up to {$Re_\theta=2500$} studied through simulation and experiment.
\newblock Physics of Fluids \textbf{21}(5), 051702 (2009).
\newblock \doi{{10.1063/1.3139294}}

\bibitem{ShengKatz:2008}
Sheng, J., Malkiel, E., Katz, J.: Using digital holographic microscopy for
  simultaneous measurements of 3d near wall velocity and wall shear stress in a
  turbulent boundary layer.
\newblock Experiments in Fluids \textbf{45}(6), 1023--1035 (2008).
\newblock \doi{10.1007/s00348-008-0524-2}

\bibitem{SpalartColeman:1997}
Spalart, P., Coleman, G.: Numerical study of a separation bubble with heat
  transfer.
\newblock European Journal of Mechanics B - Fluids \textbf{16}(2), 169--189
  (1997)

\bibitem{Willert:2015}
Willert, C.E.: High-speed particle image velocimetry for the efficient
  measurement of turbulence statistics.
\newblock Experiments in Fluids \textbf{56}(1), 17 (2015).
\newblock \doi{10.1007/s00348-014-1892-4}

\end{thebibliography}

\end{document}